\documentclass[prd,12pt,tightenlines,nofootinbib]{revtex4}
\usepackage{amssymb,amsthm}
\usepackage{amsmath,amssymb}
\newcommand{\be}{\begin{equation}}
\newcommand{\ee}{\end{equation}}
\newcommand{\bea}{\begin{eqnarray}}
\newcommand{\eea}{\end{eqnarray}}

\begin{document}

\title{Electrodynamics in a Background Chiral Field}

\author{F.T. Brandt$^{a}$}\email{fbrandt@usp.br}
\author{D.G.C. McKeon$^{b,c}$}\email{dgmckeo2@uwo.ca}
\author{A. Patrushev$^{b}$}\email{apatrus@uwo.ca}
\affiliation{$^a$Instituto de F\'\i sica,Universidade de S\~ao Paulo, S\~ao Paulo, SP 05315-970 Brazil}
\affiliation{$^b$Department of Applied Mathematics, University of Western Ontario, London Canada N6A 5B7}
\affiliation{$^c$Department of Mathematics and Computer Science, Algoma University, Sault St.Marie Canada P6A 2G4, Canada}

\begin{abstract}

We consider the one loop effective action in four dimensional Euclidean space for a background chiral field coupled to a spinor field.
It proves possible to find an exact expression for this action if the mass $m$ of the spinor vanishes.
If $m$ does not vanish, one can make a perturbative expansion in powers of the axial field that contributes to the chiral field,
while treating the contribution of the vector field exactly when it is a constant.
The analogous problem in two dimensions is also discussed.
\end{abstract}

\maketitle

\section{Introduction}

Parity violating interactions with a spinor field yield several interesting consequences, among them an anomalous
divergence in the axial current   \cite{Schwinger:1951nm,Adler:1969gk,Bell:1969ts} 
and the absence of bound states in a ``Coulomb'' axial potential \cite{Macdonald:1999,mckeon:2002}.  
In this paper we consider the one loop effective action for a spinor field in the presence of a constant background chiral vector field.
The analogous situation in which the interaction is parity conserving is well
known \cite{Schwinger:1951nm,Heisenberg:1935qt,weisskopf:1936,Dunne:2004nc}. 

\section{Effective Action}

If a spinor $\psi$ is in the presence of a background vector field $V^\mu$
and a background axial field $A^\mu$ in four dimensional Euclidean space we have the Lagrangian
\begin{equation}\label{eq1}
\mathcal{L} = \psi^\dagger \left[ \left(\not\!p - \not\!\!W_+P_+ - \not\!\!W_-P_- \right)-m\right]\psi
\end{equation}
where $p = -i\partial$ and $W_{\pm} = V \pm A$ are chiral fields.  (The notation used is in the appendix.)
The effective action is then given by the one loop expression
\begin{equation}\label{eq2}
\Gamma_4 = \ln\det \left(\not\!p - \not\!\!W_+P_+ - \not\!\!W_-P_- - m\right).
\end{equation}
We now rewrite Eq. \eqref{eq2} as
\begin{equation}\label{eq3}
\Gamma_4 = \left[ \ln\det \left(\not\!p - \not\!\!W_+P_+ - \not\!\!W_-P_- \right) + \ln\det \left( 1 -
\frac{m}{\not\!p - \not\!\!W_+P_+ - \not\!\!W_-P_- }\right)\right]
\end{equation}
and then expand the second term in Eq. \eqref{eq3} so that
\begin{equation}\label{eq4}
\ln\det \left( 1 - \frac{m}{\not\!p - \not\!\!W_+P_+ - \not\!\!W_-P_- }\right)
=-{\rm  tr}   \sum_{n=1}^\infty \,\frac{1}{n} \left( \frac{m}{\not\!p - \not\!\!W_+P_+ - \not\!\!W_-P_- }\right)^n.
\end{equation}
We now rewrite
\begin{align}\label{eq5}
\frac{1}{\not\!p - \not\!\!W_+P_+ - \not\!\!W_-P_- }
&= \frac{1}{\not\!p}  \;\frac{1}{1-\frac{1}{\not p}\left(\not\!\!W_+P_+ + \not\!\!W_-P_-\right)}\nonumber \\
&= \frac{1}{\not\!p}\sum_{n=0}^\infty \left[ \frac{1}{\not\!p}\left(\not\!\!W_+P_+ + \not\!\!W_-P_-\right)\right]^n\nonumber\\
\intertext{\rm{which by the properties of the projection operators $P_\pm$ becomes}}
&= \frac{1}{\not\!p} \sum_{n=0}^\infty \left[ \left( \frac{1}{\not\!p} \not\!\!W_+\right)^n P_+ +
\left( \frac{1}{\not\!p} \not\!\!W_-\right)^n P_-\right]\nonumber \\
&= \frac{1}{\not\!p - \not\!\!W_+} P_++  \frac{1}{\not\!p - \not\!\!W_-} P_- \,.
\end{align}
Similarly, we have for the first term in Eq. \eqref{eq3}
\begin{align}\label{eq6}
\ln\det \left(\not\!p - \not\!\!W_+P_+ - \not\!\!W_-P_-\right)&= {\rm tr}\left[\ln \not\!p - \sum_{n=1}^\infty \frac{1}{n}
\left(\frac{1}{\not\!p} \not\!\!W_+P_+ + \frac{1}{\not\!p} \not\!\!W_-P_-\right)^n\right]\nonumber\\
&= {\rm tr}\left[\left(\ln ( \not\!p - \not\!\!W_+)\right)P_+ + \left(\ln ( \not\!p - \not\!\!W_-)\right)P_-\right].
\end{align}
Together, Eqs. \eqref{eq3}, \eqref{eq4}, \eqref{eq5} and \eqref{eq6} show that
\begin{align}\label{eq7}
\Gamma_4 &= {\rm tr}\left[(\ln \not\!\Pi_+)P_+ + (\ln \not\!\Pi_-)P_- - \frac{m}{1}\left( \frac{1}{\not\!\Pi_+} P_+ +
\frac{1}{\not\!\Pi_-} P_-\right)   \right.\nonumber \\
 &\quad- \frac{m^2}{2}\left( \frac{1}{\not\!\Pi_-} \frac{1}{\not\!\Pi_+}P_+ +
\frac{1}{\not\!\Pi_+} \frac{1}{\not\!\Pi_-} P_-\right)\nonumber \\
 &\quad\quad- \frac{m^3}{3}\left( \frac{1}{\not\!\Pi_+} \frac{1}{\not\!\Pi_-} \frac{1}{\not\!\Pi_+}P_+ +
\frac{1}{\not\!\Pi_-}\frac{1}{\not\!\Pi_+} \frac{1}{\not\!\Pi_-} P_-\right)\nonumber \\
 &\qquad\qquad -   \ldots\Bigg]
\end{align}
where $\Pi_\pm \equiv p - W_\pm$.

If we now use the identity
\begin{equation}\label{eq8}
{\rm tr}X = \frac{1}{2} {\rm tr}\left[X + \gamma^5 X \gamma^5\right]
\end{equation}
then we see that terms in Eq. \eqref{eq7} with odd powers of $m$ vanish. This reduces Eq. \eqref{eq7} to
\begin{equation}\label{eq9}
\Gamma_4 =\frac{1}{2} {\rm tr}\left\{ \left[\ln\left( \not\!\Pi_+^2 \left( 1 - \frac{m^2}{\not\!\Pi_-\not\!\Pi_+}\right)\right)\right] P_+ +
\left[\ln \left(\not\!\Pi^2_- \left(1 - \frac{m^2}{\not\!\Pi_+\not\!\Pi_-}\right)\right)\right]P_-\right\}.
\end{equation}
Under ``charge conjugation'' we find that
\begin{align}\label{eq10}
C^{-1} & \left(\not\!p - \not\!\!W_+P_+ - \not\!\!W_-P_- - m  \right)C\nonumber\\
&= \left[\not\!p + \not\!\!W_+P_- + \not\!\!W_-P_+ - m  \right]^T
\end{align}
and so Eq. \eqref{eq2}
is symmetric under the replacement $W_\pm \rightarrow - W_\mp$.  (In ref. \cite{Maroto:1998zc} 
the fact that $p^{\mu T} = -p^\mu$ was  ignored.)

\section{Explicit Evaluation of the Effective Action}

Evaluation of $\Gamma$ in Eq. \eqref{eq9} in closed form when
$m^2 \neq 0$ involves having to determine ${\rm tr}\ln(\not\!\Pi_\pm \not\!\Pi_\mp - m^2)$.
If $\not\!\!W_\pm \neq  \not\!\!W_\mp$ this is prohibitively
difficult,  even if $W_\pm = \pm A$.  In this case we must consider
\begin{equation}\label{eq11}
{\rm tr}\ln\left[(\not\!p \pm \not\!\!A)(\not\!p \mp \not\!\!A) - m^2\right] = {\rm tr}\ln \left[\left(p^\mu \mp i\sigma^{\mu\nu}A^\nu\right)^2 + 2A^2
\pm iA^{\lambda}_ {,\lambda} - m^2   \right]
\end{equation}
which, though it is well suited for a perturbative expansion in powers of $A^\mu$
\cite{McKeon:1993fq,Dilkes:1998qt}, 
does not lend itself to being evaluated even
when $A^\mu$ corresponds to there being a constant field strength.

However, if $m^2 = 0$, or if Eq. \eqref{eq9} were expanded to some finite order in powers of $m^2$, then one is faced with evaluation of only
$\frac{1}{2} (\Lambda_+ + \Lambda_-)$ where $\Lambda_\pm = {\rm tr}\left[\ln \not\!\Pi_\pm^2\right]P_\pm$.
In refs.  \cite{Schwinger:1951nm,Heisenberg:1935qt,weisskopf:1936}, 
it is shown that since
$(\not\!p - \not\!V)^2 = (p^\mu - V^\mu)^2 - \frac{1}{2}\sigma^{\mu\nu}F^{\mu\nu}\,\, (F = \partial \wedge V)$ the gamma matrix trace
occurring in $\Lambda_\pm$ involves
\begin{align}\label{eq12}
{\rm tr}\,e^{\frac{1}{2}F^{\mu\nu}\sigma^{\mu\nu}t} P_\pm &= {\rm tr}\Bigg\{\cosh K_-P_+ + \cosh K_+ P_- \nonumber\\
&\left.+ \frac{t}{2} \sigma^{\mu\nu} F^{\mu\nu} \left(\frac{\sinh K_-}{K_-} P_+ + \frac{\sinh K_+}{K_+} P_-  \right)\right\}P_\pm\nonumber\\
&= 4 \cosh K_\mp
\end{align}
where $K_\pm^2 = \frac{t^2}{2} \left[F^{\mu\nu}F^{\mu\nu} \pm F^{\mu\nu}F^{*\mu\nu}\right]$.
We thus see that the presence of the
chiral projection operator $P_\pm$ in Eq. \eqref{eq9} serves to eliminate the
contribution of $\cosh K_\pm$  as well as $\sinh K_+$ and $\sinh K_-$, leaving only $4 \cosh K_\mp$.

The background field strength $W_\pm$ in the gauge $x \cdot W_\pm = 0$ can be expanded
in powers of the field strength $F_\pm$   \cite{Cronstrom:1980hj,Leupold:1996hx,Leupold:1996cx}, 
\begin{equation}\label{eq13}
W_\pm^\mu = \sum_{n=0}^\infty \frac{-1}{n!(n+2)} x^\nu x^{\lambda_{1}} \ldots x^{\lambda_{n}}\; F_\pm^{\mu\nu ,\lambda_{1}\ldots\lambda_{n}}(0).
\end{equation}
The first term in Eq. \eqref{eq13} corresponds to a constant background field as discussed in
refs. \cite{Schwinger:1951nm,Heisenberg:1935qt,weisskopf:1936};  
higher contributions are dealt with in
refs. \cite{Dunne:2004nc,Gusynin:1995bc,Lee:1989vh,Hauknes:1983xx}.  
Other special background field configurations have been considered
\cite{Schwinger:1951nm,Dunne:2004nc,Kim:2000un,Narozhnyi:1970uv,Dunne:1998ni}. 

If $m^2 = 0$ and $W_\pm = \pm A$, then we have a purely axial coupling and
\begin{equation}\label{eq14}
\Gamma_A^{(0)} = \frac{1}{2} {\rm tr}\left[\left(\ln (\not\!p - \not\!\!A)^2 \right) P_+ + \left(\ln (\not\!p + \not\!\!A)^2 \right) P_-   \right].
\end{equation}
If $A^\mu$ is in the gauge $x \cdot A = 0$ so that it is expressed in the form
of Eq. \eqref{eq13} then gauge invariance is manifestly preserved since $A^\mu$
is expressed in terms of the field strength.  If we then expand $\Gamma_A^{(0)}$
with this background field using the
Schwinger expansion as in ref.  \cite{Schwinger:1951nm,McKeon:1986rc}, 
then the three point function $\left\langle  AAA \right\rangle$
vanishes.  However, again  computing $\left\langle  AAA \right\rangle$ but with plane wave
background axial fields, the three point function is consistent with the axial anomaly
 \cite{Schwinger:1951nm,Adler:1969gk,Bell:1969ts}. 

If $m^2 \neq 0$ when $W_\pm = \pm A$ then Eq. \eqref{eq9} reduces to
\begin{align}\label{eq15}
\Gamma_A &=\frac{1}{2}  {\rm tr} \Bigg\{ \left[\ln \left( (\not\!p + \not\!\!A)(\not\!p - \not\!\!A) - m^2\right)\right]P_+ +
\left[ \ln \left( (\not\!p - \not\!\!A)(\not\!p + \not\!\!A) - m^2\right)\right]P_- \nonumber\\
  &\qquad\quad+\left. \frac{1}{2} \left[ \ln (\not\!p - \not\!\!A)^2 - \ln(\not\!p + \not\!\!A)^2\right]\gamma_5\right\}.
\end{align}
There doesn't appear to be a way of evaluating this in closed form when even $A^\mu = - \frac{1}{2} F^{\mu\nu} x^\nu$ if $m^2\neq0$, though with this background field
$\left\langle  AAA \right\rangle = 0$.  With a plane wave background field the axial anomaly can however be recovered
\cite{McKeon:1986rc}  
when $\left\langle  AAA \right\rangle$ is
computed by applying the Schwinger expansion \cite{Schwinger:1951nm}  
to Eq. \eqref{eq15}.

Although it doesn't appear to be feasible to compute $\Gamma_4$ when there is a
constant strength $\partial^\mu A^\nu - \partial^\nu A^\mu$ in Eq. \eqref{eq1}, we can consider the case in which $\Gamma_4$ is restricted
to being linear in the external axial
field and the vector field is taken to be constant.  In this case we begin by using Eq. \eqref{eq8} to write
\begin{equation}\label{eq16}
\Gamma_4 = \frac{1}{2} \ln \det \left[(\not\!p - \not\!V - \not\!\!A\gamma^5)^2 - m^2\right].
\end{equation}
Dropping those terms in Eq. \eqref{eq12} that cannot contribute to the
contribution to $\Gamma_4$ that are linear in $A_\mu$, we see that
upon letting $m^2 \rightarrow -m^2$,
\begin{align}\label{eq17}
\Gamma_4 \approx \frac{1}{2} \ln \det & \left[ (p-V)^2 + m^2 - \frac{1}{2} F^{\mu\nu} \sigma^{\mu\nu} + iA^{\mu , \mu} \gamma^5 \right. \\
& \qquad \quad \left. + i\sigma^{\mu\nu}\left(2A^\mu p^\nu + \frac{i}{2} G^{\mu\nu} - 2A^\mu V^\nu\right)\gamma^5\right]\nonumber
\end{align}
where $F^{\mu\nu} = \partial^\mu V^\nu - \partial^\nu V^\mu$ and $G^{\mu\nu} = \partial^\mu A^\nu - \partial^\nu A^\mu$.  If we
now employ operator regularization to expand $\Gamma_4$ in Eq. \eqref{eq17}
to the term linear in $A_\mu$, we need the equations
\cite{McKeon:1986rc} 
\begin{align}\label{eq18}
\frac{1}{2} \ln \det & (H_0 + H_1) = \left. -\frac{1}{2} \frac{d}{ds}\right|_0 {\rm tr}\frac{1}{\Gamma(s)} \int_0^\infty dt\,t^{s-1} \;e^{-(H_0+H_1)t} \nonumber \\
& \left. = - \frac{1}{2} \frac{d}{ds} \right|_0 \frac{1}{\Gamma(s)} \int_0^\infty dt\;t^{s-1}\;{\rm tr}\left[e^{-H_0t} +  \frac{(-t)}{1} \;
e^{-H_0t}H_1 \right. \\
& \qquad\qquad   \left. + \frac{(-t)^2}{2} \int_0^1 du\; e^{-(1-u)H_0t} H_1 e^{-uH_0t} H_1 + \ldots\right].\nonumber
\end{align}
Upon using Eq. \eqref{eq18}, Eq. \eqref{eq17} reduces to
\begin{align}\label{eq19}
\left. \Gamma_4 \approx \frac{1}{2} \frac{d}{ds}\right|_0  \frac{1}{\Gamma(s)}  & \int_0^\infty dt\,t^s \;{\rm tr}\;e^{-[(p-V)^2+m^2-\frac{1}{2} F^{\mu\nu}\sigma^{\mu\nu}]t}
\bigg[ iA_{,\mu}^\mu  \nonumber \\
& \quad \left. + i\sigma^{\lambda\sigma} \left(2A^\lambda p^\sigma + \frac{i}{2}\; G^{\lambda\sigma} - 2A^\lambda V^\sigma\right)\right]\gamma^5\; .
\end{align}
If $F^{\mu\nu}$ is constant, then by Eqs. \eqref{a1} and \eqref{a2} this becomes
\begin{align}\label{eq20}
= \frac{1}{2} \left.  \frac{d}{ds}\right|_0\; & \frac{1}{\Gamma(s)}\int_0^\infty dt\,t^s \; {\rm tr}\;e^{[(p-V)^2+m^2]t} \bigg[ (\cosh K_-)P_+ + (\cosh K_+)P_-   \nonumber\\
& \left. + \left(\frac{\sinh K_-}{K_-} P_+ + \frac{\sinh K_+}{K_+} P_-\right)w^{\mu\nu}\sigma^{\mu\nu}\right]\nonumber \\
& \left[iA_{, \lambda}^\lambda + i\sigma^{\lambda\sigma}\left(2A^\lambda p^\sigma + \frac{i}{2} G^{\lambda\sigma} - 2A^\lambda V^\sigma\right)   \right] \gamma_5
\end{align}
where $w^{\mu\nu} = \frac{1}{2} F^{\mu\nu} t$ and $K_\pm^2 = 2(w^{\alpha\beta}w^{\alpha\beta} \pm w^{\ast\alpha\beta} w^{\alpha\beta})$.

Evaluating the $\gamma$-matrix traces in Eq. \eqref{eq20} leads to
\begin{align}\label{eq21}
&= \left. \frac{d}{ds}\right|_0   \frac{i}{\Gamma(s)} \int_0^\infty dt\; t^s \; {\rm tr}\; e^{-[(p-V)^2 + m^2]t} \left\lbrace (\cosh K_- - \cosh K_+)A_{,\lambda}^\lambda \right. \nonumber\\
&\qquad + 2\left[\left(\frac{\sinh K_-}{K_-} - \frac{\sinh K_+}{K_+}\right) w^{\lambda\sigma} - 2 \left( \frac{\sinh K_-}{K_-}   + \frac{\sinh K_+}{K_+}\right)
w^{\ast\lambda\sigma}  \right]\nonumber\\
& \qquad \left. \left[2A^\lambda p^\sigma + \frac{i}{2} G^{\lambda\sigma} - 2 A^\lambda V^\sigma\right] \right\rbrace .
\end{align}
When $V^\mu = -\frac{1}{2} F^{\mu\nu} x^\nu$, then the result of Schwinger \cite{Schwinger:1951nm}  
\begin{align}\label{eq22}
\langle   x|e^{-(p-V)^2t} |y \rangle  &= \frac{i}{(4\pi t)^2} \exp \left(i\int_y^x dz \cdot V(z)  \right)e^{-L(t)}\\
\qquad \qquad\qquad & \exp\left( - \frac{1}{4} (x-y) \cdot F \cdot \cot (Ft) \cdot (x-y)\right)\nonumber
\end{align}
can be used to compute the functional trace in Eq. \eqref{eq21}.
(Here we have $L(t) = \frac{1}{2} {\rm tr}\ln ((Ft)^{-1} \sin (Ft))$.)  In particular, it follows
from Eq. \eqref{eq22} that
\begin{align}\label{eq23}
{\rm tr}\; e^{-(p-V)^2t} A^\lambda p^\sigma  &= {\rm tr}\int dz \langle x|e^{-(p-V)^2t} |z \rangle  i\partial_y^\sigma \langle  z|A^\sigma |y \rangle \nonumber \\
&= \int dx \int dy \; \delta(x-y) i\partial_y^\sigma \bigg[ \frac{i}{(4\pi t)^2} \exp \left( i\int_y^x dz V(z) \right)e^{-L(t)}  \nonumber \\
& \qquad  \exp\left( - \frac{1}{4} (x-y) \cdot F \cdot \cot (Ft) \cdot (x-y)\right)A^\lambda (y)\bigg]\nonumber\\
& = \frac{i}{(4\pi t)^2} \,e^{-L(t)} \int dx \left[V^\sigma (x)  A^\lambda(x) + i\partial_x^\sigma A^\lambda(x)   \right].
\end{align}
Substitution Eqs. \eqref{eq22} and \eqref{eq23} into Eq. \eqref{eq21} leads to
\begin{align}\label{eq24}
\Gamma_4 & \approx \frac{-1}{(4\pi)^2} \left. \frac{d}{ds}\right|_0\frac{1}{\Gamma(s)} \int_0^\infty dt\; t^{s-2} e^{-L(t)-m^2t}
\int dx \bigg\{(\cosh K_- - \cosh K_+)A^{\mu ,\mu}(x)  \nonumber \\
& \left. - \frac{i}{2} G^{\lambda\sigma}(x)t\left[\left(\frac{\sinh K_-}{K_-} - \frac{\sinh K_+}{K_+}  \right) F^{\lambda\sigma} -
 \left(\frac{\sinh K_-}{K_-} + \frac{\sinh K_+}{K_+}  \right) F^{\ast\lambda\sigma} \right]\right\rbrace .
\end{align}
Expanding Eq. \eqref{eq24} to lowest order in $F^{\lambda\sigma}$ results in
\begin{align}\label{eq25}
\Gamma_4 \approx \frac{1}{(4\pi)^2} & \left. \frac{d}{ds}\right|_0 \frac{1}{\Gamma(s)}\int_0^\infty dt\; t^{s-2} e^{-m^2t} \int dx\left[\frac{1}{2} t^2 F^{\lambda\sigma}F^{\ast\lambda\sigma}
A^{\mu , \mu} (x)  \right.\nonumber \\
& - it\, G^{\lambda\sigma}(x)F^{\ast\lambda\sigma}\bigg]\nonumber \\
&= \frac{1}{(4\pi)^2} \int dx \left[\frac{1}{m^2} F^{\lambda\sigma} F^{\ast\lambda\sigma} A^{\mu , \mu} + i(\ln m^2) G^{\lambda\sigma}  F^{\ast\lambda\sigma}  \right].
\end{align}
Neither term in Eq. \eqref{eq25} would arise from the calculation of
one-loop Feynman diagrams with plane wave external fields.  For $F_{\mu\nu}$
being a constant field, the first term in Eq. \eqref{eq25} is a total derivative.
When either $F$ or $G$ (or both) are non-constant the second term is also a total derivative.

\section{The Two-dimensional Limit}

The two dimensional limit of massive electrodynamics has been considered in refs.
\cite{Adam:1997hs,Krasnansky:2006sx}. 
If there is an axial coupling between the spinor and an external
axial field, this leads to the one-loop effective action
\begin{equation}\label{eq26}
\Gamma_2 = \ln\det (\not\!p - \not\!\!A \sigma^3 - m)\qquad (p \equiv -i\partial) .
\end{equation}
However, as $\gamma^\mu\sigma^3 = \epsilon^{\mu\nu}\gamma_\nu$, this becomes
\begin{equation}\label{eq27}
\Gamma_2 = \ln\det (\not\!p - A_\mu \epsilon^{\mu\nu}\gamma_\nu - m).
\end{equation}
Consequently, if the background field $A_\mu$ corresponds to a constant
field strength $A_\mu = - \frac{1}{2} F_{\mu\nu}x^\nu = - \frac{f}{2} \epsilon_{\mu\nu}x^\nu$,
then Eq. \eqref{eq27} reduces to
\begin{equation}\label{eq28}
\Gamma_2 = \ln\det\left(\not\!p - \frac{f}{2} \not\!x - m\right)
\end{equation}
which is what would be obtained if there were a parity conserving coupling with an
external vector field $V_\mu = \frac{1}{4} f\partial_\mu(x^2)$ which
corresponds to a pure gauge field.  This effective action should thus be independent of $f$,
which we will show explicitly by using Schwinger's technique \cite{Schwinger:1951nm}. 

If now
\begin{equation}\label{eq29}
\Pi_\mu = p_\mu - \frac{f}{2} x_\mu
\end{equation}
then Eq. \eqref{eq28} becomes
\begin{equation}\label{eq30}
\Gamma_2 = \ln {\det}^{1/2} (\not\!\Pi + m)(\not\!\Pi - m) = \frac{1}{2} \ln\det (\Pi^2 - m^2)
\end{equation}
upon using the two dimensional analogue of Eq. \eqref{eq8} and
\begin{equation}\label{eq31}
\left[ \Pi_\mu , \Pi_\nu \right] = 0.
\end{equation}
Regulating $\Gamma_2$ using the $\zeta$-function \cite{Salam:1974xe,Hawking:1976ja} 
we have
\begin{equation}\label{eq32}
\left. \Gamma_2 = -\frac{1}{2} \frac{d}{ds}\right|_0 \frac{1}{\Gamma(s)} {\rm tr}\int_0^\infty d\,it (it)^{s-1} e^{i(m^2-\Pi^2)t}.
\end{equation}
To evaluate the functional trace in Eq. \eqref{eq32}, we use the Hamiltonian
approach of ref. \cite{Schwinger:1951nm}, 
defining
\begin{equation}\label{eq33}
\langle x(t)|y(0)\rangle =\langle x|e^{-iH  t}|y  \rangle
\end{equation}
with
\begin{equation}\label{eq34}
H = -\Pi^2\, .
\end{equation}
The equations
\begin{subequations}\label{eq35}
\begin{eqnarray}
i\frac{\partial\Pi^\mu(t)}{\partial t} &=& \left[\Pi^\mu (t),H\right]\\
i\frac{\partial x^\mu}{\partial t} &=& \left[x^\mu (t),H\right]
\end{eqnarray}
\end{subequations}
can be integrated to give
\begin{subequations}\label{eq36}
\begin{eqnarray}
\Pi^\mu(t) &=& \Pi^\mu (0)\\
x^\mu (t) &=& -2\Pi_\mu (0).
\end{eqnarray}
\end{subequations}

Since Eq. \eqref{eq36} is identical to the equations
that arise if $f = 0$, we see that the effective action in two dimensions
for a spinor in the presence of a constant background axial field is just that of a free field.

\section{Conclusions}

We thus see that the one-loop effective action for a spinor in the presence of
a constant background chiral field is closely related to that of considered
in refs. \cite{Schwinger:1951nm,Heisenberg:1935qt,weisskopf:1936,Dunne:2004nc}  
provided $m^2 = 0$.  The case in which $m^2 \neq 0$ in four dimensions
has not as yet been given in closed form.  Higher order calculations, or those
involving non-constant background fields are currently being considered,
as is that all-orders approach in the presence of a
weak background field \cite{Affleck:1981bma,Dunne:2005sx}.  

We note the use of projection operators in conjunction with background gauge fields in ref. \cite{Hur:2010}.

\bigskip

\noindent
{\Large\bf{Acknowledgments}}\\
We would like to thank Christian Schubert for helpful discussions. Also, Roger Macleod had a helpful suggestion.
We appreciate the valuable remarks of the anonymous referee.
F. T. Brandt would like to thank CNPq for financial support and the
hospitality of the Department of Applied Mathematics of the University
of Western Ontario, Canada.

\noindent

\appendix*

\section{}

In four dimensional Euclidean space we have the conventions
\[\left\{ \gamma^\mu , \gamma^\nu \right\} =
2\delta^{\mu\nu}, \quad \quad \left[ \gamma^\mu , \gamma^\nu \right] = 2i\sigma^{\mu\nu}\nonumber\]

\[\left[ \sigma^{\mu\nu} , \sigma^{\lambda\sigma} \right]=
2i\left(\delta^{\mu\lambda} \sigma^{\nu\sigma} - \delta^{\mu\sigma} \sigma^{\nu\lambda} +
        \delta^{\nu\sigma} \sigma^{\mu\lambda}  - \delta^{\nu\lambda}   \sigma^{\mu\sigma}\right)\nonumber\]

\[\left\{\sigma^{\mu\nu}, \sigma^{\lambda\sigma}\right\} =
2 \left(\delta^{\mu\lambda} \delta^{\nu\sigma} - \delta^{\mu\sigma}\delta^{\nu\lambda}\right)
- 2\epsilon^{\mu\nu\lambda\sigma}\gamma^5\nonumber\]

\[\gamma^\alpha\gamma^\beta \gamma^\lambda =
\delta^{\alpha\beta}\gamma^\lambda - \delta^{\alpha\lambda}\gamma^\beta +
\delta^{\beta\lambda}\gamma^\alpha - \epsilon^{\alpha\beta\lambda\rho}\gamma^\rho\gamma^5\nonumber\]

\[\epsilon^{1234} = 1, \qquad  \gamma^5 = \gamma^1\gamma^2\gamma^3\gamma^4 , \qquad {\rm tr}\;\gamma^5 = 0\nonumber \]

\[\sigma^{\mu\nu}\gamma^5 = \epsilon^{\mu\nu\lambda\sigma}\sigma^{\lambda\sigma}.\nonumber\]

\[ P_\pm = \frac{1\pm\gamma^5}{2}\quad , (P_\pm)^2 = P_\pm\quad ,\quad P_\pm P_\mp = 0\nonumber\]

\[\qquad\qquad P_\pm\gamma^\mu = \gamma^\mu P_\mp\quad , \quad P_\pm\gamma^5 =  \gamma^5 P_\pm .\]
These show that if
\be\label{a1}
e^{\lambda w^{\mu\nu}\sigma^{\mu\nu}} = (A_+(\lambda)P_+ + A_-(\lambda)P_-) + (B_+(\lambda)P_+ + B_-(\lambda)P_-)w^{\mu\nu}\sigma^{\mu\nu}
\ee
then the differential equation
\begin{equation}
\frac{d}{d\lambda} e^{\lambda w^{\mu\nu}\sigma^{\mu\nu}} =  w^{\mu\nu}\sigma^{\mu\nu} e^{\lambda \sigma^{\mu\nu}w^{\mu\nu}}\nonumber
\end{equation}
leads to
\begin{equation}
\dot{A}_\pm = K_\mp^2 B_\pm\;,\qquad  \dot{B}_\pm = A_\pm \qquad(A_\pm(0) = 1, \; B_\pm (0)= 0)\nonumber
\end{equation}
where $K_\pm^2 = 2(w^{\mu\nu}w^{\mu\nu} \pm w^{\mu\nu}w^{*\mu\nu})$
and $w^{*\mu\nu} = \frac{1}{2}\epsilon^{\mu\nu\lambda\sigma}w^{\lambda\sigma}$.  These have the solution when $\lambda = 1$
\be\label{a2}
A_\pm = \cosh K_\mp\qquad\quad B_\pm = \frac{\sinh K_\mp}{K_\mp} .
\ee
The ``charge conjugation'' matrix $C$ satisfies
$C^{-1}\gamma^\mu C = -\gamma^{\mu T}, C^{-1}\gamma^5 C = \gamma^{5T}$.

In two dimensional Minkowski space, we take
\begin{align}
&g^{00} = 1 = -g^{11}\quad {\rm{and}}
\quad \gamma^0 = \sigma^1\;,\gamma^1 = i\sigma^2\quad {\rm{so\; that}}\nonumber\\
&{\rm{if}}\quad \epsilon^{01} = 1 = \epsilon_{10},\; {\rm{then}}
\quad \gamma^\mu\gamma^\nu = g^{\mu\nu} - \epsilon^{\mu\nu}\sigma^3 \;\; {\rm{and}} \;\;
\gamma^\mu \sigma^3 = \epsilon^{\mu\nu}\gamma_\nu\nonumber
\end{align}
(where $\sigma^i$ is a Pauli spin matrix).


\end{document}